\documentclass[aps,prd,twocolumn]{revtex4-2}
\usepackage{graphicx}
\usepackage{amssymb}
\usepackage{amsmath}
\usepackage{float}

\begin{document}

\title{Magnification of Classical Multimessenger Signals in Stationary and Axisymmetric Spacetimes with Separable Equations of Motion}

\author{Torben C. Frost}

\affiliation{Kavli Institute for Astronomy and Astrophysics, Peking University, 100871 Beijing, China.\\
e-mail: torben.frost@pku.edu.cn}

\date{June 30th, 2025}

\begin{abstract}
When photons, gravitational waves, and massive particles such as neutrinos are gravitationally lensed the signals detected by telescopes or detectors on and around Earth are usually either magnified or demagnified. However, for stationary and axisymmetric spacetimes conventional methods for calculating the magnification factor usually only allow to calculate it relative to the spacetime position of the source but not with respect to the source's reference frame. While this may be sufficient when we want to investigate the properties of the lens, when we want to investigate the source and its properties we need to relate the detected signals to the same signals in the reference frame of their source. In this paper we will now show that for stationary and axisymmetric spacetimes which possess a Carter constant, we can use the constants of motion and the tetrad formalism to derive a magnification factor which relates a signal detected in the reference frame of an observer to the same signal in the reference frame of its source.
\end{abstract}
\maketitle
\section{Introduction}
When we observe sources around an astrophysical object which can be described by a stationary and axisymmetric spacetime, independent of the fact whether these sources are accretion disks, regular stars, neutron stars, or compact object binaries consisting of neutron stars and black holes, electromagnetic radiation, gravitational waves, and neutrinos emitted by these sources are gravitationally lensed and beams consisting of these messengers are either focussed or defocussed. As a consequence, signals observed by telescopes and detectors on and orbiting Earth are magnified or demagnified. 

In many cases, already in the weak-field regime determining the exact magnification of these signals is an extremely difficult task since even when we observe multiple signals of the same messenger we can only determine the relative magnification between them. This gets even more problematic when we investigate gravitational lensing in a curved spacetime. As pointed out in the living review of Perlick \cite{Perlick2004b} in a curved spacetime even the relative magnification cannot be directly observed since in general the signals travelled different affine distances. In addition, when we only observe one signal, or two signals from different messengers, e.g., electromagnetic radiation and neutrinos, determining the magnification factor becomes even more difficult. In particular, it is very challenging to calculate the original flux or intensity in the local coordinate system of the source. However, for the correct interpretation of observed messenger signals as well as to gain a better understanding of their sources, being able to determine the flux or the intensity in the reference frame of the source from an observed signal is of crucial importance. 

While in the weak-field regime one can simply calculate the magnification factor using the Jacobian of the lens equation as, e.g., described in the book of Petters, Levine, and Wambsganss \cite{Petters2001}, when we use the full formalism of general relativity we can define the magnification factor in different ways. Here, in all of them writing down an exact lens equation, following Frittelli and Newman \cite{Frittelli1999}, and thus solving the equations of motion plays an important role. When we write down the exact lens equation we can transfer many of the concepts from the weak-field formalism to the exact approach, see, e.g., the work of Frittelli, Kling, and Newman \cite{Frittelli2001a,Frittelli2001b}. In particular we see that also in the exact formalism the magnification factor can be calculated from the determinant of the Jacobi matrix of the exact lens equation, see, e.g., the work of Kraniotis \cite{Kraniotis2011}. However, the calculated magnification factor usually compares a signal detected in a curved background to the corresponding signal in a flat background. In addition, as demonstrated, e.g., in the living review of Perlick \cite{Perlick2004b}, one can also use the affine and the shape parameters, to define the magnification factor. However, in this case it compares the size of a source on the celestial sphere of an observer at an affine distance $s$ in a curved spacetime to the apparent size of the source on an observer's celestial sphere at the same affine distance in the Minkowski spacetime. However, as pointed out by Perlick \cite{Perlick2004a} commonly we do not have a flat background which makes using these definitions for astrophysical observations rather difficult. In addition, in both cases it is rather difficult to directly calculate the magnification between the source frame and the observer frame. 

In this paper we will now show that, when the line element of the spacetime is stationary and axisymmetric, and the corresponding equations of motion are separable and can be written down in an analytic form, it is possible to use the constants of motion along lightlike and timelike geodesics (including electromagnetic radiation in a cold, non-magnetised, and pressureless plasma; hereafter for simplicity only plasma) to obtain a relation between the celestial coordinates in the local coordinate frames of a source and an observer. This relation can then be used to calculate the magnification of the messenger signals between the source frame and the observer frame.

For this purpose in Section~\ref{Sec:Sep} we will first briefly review the separability conditions for the equations of motion of stationary and axisymmetric spacetimes in general relativity following Bezd\v{e}kov\'{a}, Perlick, and Bi\v{c}\'{a}k \cite{Bezdekova2022}. In Section~\ref{Sec:MFD} we will then show how one can use the constants of motion and the tetrad formalism to calculate the determinant of the Jacobi matrix and the magnification factor. In Section~\ref{Sec:ASSSS} we will then demonstrate how one can use this approach for calculating the magnification factor for a static observer and static sources in a general static and spherically symmetric spacetime for light rays, high-frequency gravitational waves, and massive particles in vacuum, and light rays in a homogeneous plasma. In Section~\ref{Sec:SO} we will then briefly summarise the results and discuss their potential application to accretion disks around supermassive black holes.

\section{Separability of the Equations of Motion}\label{Sec:Sep}
In the following section we want to show how one can calculate the magnification factor for axisymmetric and stationary spacetimes with separable equations of motion. However, since many readers may not be aware of the conditions under which the equations of motion of stationary and axisymmetric spacetimes are separable we will briefly review them in this section. 

Carter \cite{Carter1968} was the first to show that one can separate the equations of motion of the Kerr spacetime for motion along lightlike and timelike geodesics. On the other hand for the Kerr spacetime with a cold, non-magnetised, and pressureless plasma the conditions under which the equations of motion are seperable were only investigated relatively recently by Perlick and Tsupko \cite{Perlick2017}. Bezd\v{e}kov\'{a}, Perlick, and Bi\v{c}\'{a}k \cite{Bezdekova2022} then extended this investigation to a general stationary and axisymmetric spacetime. Thus, since we want to develop a method to calculate the magnification factor for a general stationary and axisymmetric spacetime with separable equations of motion for light rays, high-frequency gravitational waves, and massive particles in vacuum, and light rays in a plasma, in this paper we will closely follow the derivation presented in Bezd\v{e}kov\'{a}, Perlick, and Bi\v{c}\'{a}k \cite{Bezdekova2022}. 

Let us start with the generalised stationary and axisymmetric line element in geometric units with $c=G=1$:
\begin{eqnarray}
&g_{\mu\nu}\text{d}x^{\mu}\text{d}x^{\nu}=-G_{1}(r,\vartheta)\text{d}t^2+2G_{2}(r,\vartheta)\text{d}t\text{d}\varphi\\
&+G_{3}(r,\vartheta)\text{d}\varphi^2+G_{4}(r,\vartheta)\text{d}r^2+G_{5}(r,\vartheta)\text{d}\vartheta^2,\nonumber
\end{eqnarray}
where $G_{1}(r,\vartheta)-G_{5}(r,\vartheta)$ are for now general functions of $r$ and $\vartheta$. In the following we will also need the non-vanishing components of the inverse metric tensor. They read
\begin{eqnarray}
g^{tt}=-\frac{G_{3}(r,\vartheta)}{G_{1}(r,\vartheta)G_{3}(r,\vartheta)+G_{2}(r,\vartheta)^2},
\end{eqnarray}
\begin{eqnarray}
g^{t\varphi}=g^{\varphi t}=\frac{G_{2}(r,\vartheta)}{G_{1}(r,\vartheta)G_{3}(r,\vartheta)+G_{2}(r,\vartheta)^2},
\end{eqnarray}
\begin{eqnarray}
g^{\varphi\varphi}=\frac{G_{1}(r,\vartheta)}{G_{1}(r,\vartheta)G_{3}(r,\vartheta)+G_{2}(r,\vartheta)^2},
\end{eqnarray}
\begin{eqnarray}
g^{rr}=\frac{1}{G_{4}(r,\vartheta)},~~~g^{\vartheta\vartheta}=\frac{1}{G_{5}(r,\vartheta)}.
\end{eqnarray}
Now let us write down the Hamiltonians for light rays, high-frequency gravitational waves, and massive particles in vacuum, and the Hamiltonian for light rays in a plasma. They read
\begin{eqnarray}
\mathcal{H}_{\text{v}}=\frac{1}{2}g^{\mu\nu}p_{\mu}p_{\nu},
\end{eqnarray}
\begin{eqnarray}
\mathcal{H}_{\text{pl}}=\frac{1}{2}\left(g^{\mu\nu}p_{\mu}p_{\nu}+E_{\text{pl}}(x)^2\right),
\end{eqnarray}
where the $p_{\mu}$ are the components of the four-momentum of the messengers and $E_{\text{pl}}(x)$ is the characteristic energy of the plasma at the spacetime coordinates $x$. Note that the plasma energy is related to the plasma frequency $\omega_{\text{pl}}(x)$ by $E_{\text{pl}}(x)=\hbar\omega_{\text{pl}}(x)$. Since in this paper we only consider stationary and axisymmetric spacetimes we have two conserved quantities associated with the two Killing-vector fields $\xi_{t}=\partial_{t}$ and $\xi_{\varphi}=\partial_{\varphi}$. These are the energy $E$ and the $z$-component of the angular momentum $L_{z}$. As a consequence we have $p_{t}=-E$ and $p_{\varphi}=L_{z}$. 

Now we can derive the equations of motion using Hamilton's equations
\begin{eqnarray}\label{eq:EoMD}
\dot{x}^{\mu}=\frac{\partial\mathcal{H}}{\partial p_{\mu}}~~~\text{and}~~~\dot{p}_{\mu}=-\frac{\partial\mathcal{H}}{\partial x^{\mu}},
\end{eqnarray}
and $\mathcal{H}_{\text{v}}=\mathcal{H}_{\text{pl}}=0$ for light rays and high-frequency gravitational waves and $\mathcal{H}_{\text{v}}=-1/2$ for massive particles in vacuum. As first step we derive the equations of motion for the time coordinate $t$ and the $\varphi$ coordinate. We get
\begin{eqnarray}\label{eq:EoMt}
\dot{t}=\frac{G_{3}(r,\vartheta)E+G_{2}(r,\vartheta)L_{z}}{G_{1}(r,\vartheta)G_{3}(r,\vartheta)+G_{2}(r,\vartheta)^2},
\end{eqnarray}
\begin{eqnarray}\label{eq:EoMphi}
\dot{\varphi}=\frac{G_{1}(r,\vartheta)L_{z}-G_{2}(r,\vartheta)E}{G_{1}(r,\vartheta)G_{3}(r,\vartheta)+G_{2}(r,\vartheta)^2},
\end{eqnarray}
where we rewrote the components of the four-momentum in terms of the constants of motion. In the next step we multiply the Hamiltonians with 2, and rewrite them in the following form (note that here we already assume that the plasma term can only depend on $r$ and $\vartheta$ since otherwise the whole system would not be stationary and axisymmetric)
\begin{eqnarray}\label{eq:HamSum}
g^{\mu\nu}p_{\mu}p_{\nu}+E_{\text{p}}(r,\vartheta)^2=0,
\end{eqnarray}
where we have $E_{\text{p}}(r,\vartheta)=0$ for light rays and high-frequency gravitational waves in vacuum, $E_{\text{p}}(r,\vartheta)=1$ for massive particles in vacuum, and $E_{\text{p}}(r,\vartheta)=E_{\text{pl}}(r,\vartheta)\neq 0$ for light rays in plasma. In the next step we write down the action associated with the Hamiltonian. Here we require as constraint on the metric tensor that the action separates and we have
\begin{eqnarray}
S(t,r,\vartheta,\varphi)=S_{t}(t)+S_{r}(r)+S_{\vartheta}(\vartheta)+S_{\varphi}(\varphi).
\end{eqnarray}
Now we use that the components of the four-momentum, the action function, and the components of the four-velocity of the signal are related by
\begin{eqnarray}\label{eq:MAV}
p_{\mu}=\frac{\partial S}{\partial x^{\mu}}=g_{\mu\nu}\dot{x}^{\nu}
\end{eqnarray}
and rewrite (\ref{eq:HamSum}) as
\begin{eqnarray}
&\hspace{-0.5cm}\frac{G_{1}(r,\vartheta)L_{z}^2-2G_{2}(r,\vartheta)EL_{z}-G_{3}(r,\vartheta)E^2}{G_{1}(r,\vartheta)G_{3}(r,\vartheta)+G_{2}(r,\vartheta)^2}+\frac{1}{G_{4}(r,\vartheta)}\left(\frac{\partial S_{r}(r)}{\partial r}\right)^2\\
&+\frac{1}{G_{5}(r,\vartheta)}\left(\frac{\partial S_{\vartheta}(\vartheta)}{\partial \vartheta}\right)^2+E_{\text{p}}(r,\vartheta)^2=0.\nonumber
\end{eqnarray}
Now the only remaining freedom we have is that we can multiply with a nonzero function $F(r,\vartheta)$. Afterwards for separating the equations of motion the remaining terms have to fulfill the following conditions
\begin{eqnarray}
E_{\text{p}}(r,\vartheta)^2=\frac{E_{\text{p}r}(r)+E_{\text{p}\vartheta}(\vartheta)}{F(r,\vartheta)},
\end{eqnarray}
\begin{eqnarray}
H_{1}(r)=\frac{F(r,\vartheta)}{G_{4}(r,\vartheta)}~~~\text{and}~~~H_{2}(\vartheta)=\frac{F(r,\vartheta)}{G_{5}(r,\vartheta)},
\end{eqnarray}
and
\begin{eqnarray}\label{eq:Gi}
\frac{F(r,\vartheta)G_{i}(r,\vartheta)}{G_{1}(r,\vartheta)G_{3}(r,\vartheta)+G_{2}(r,\vartheta)^2}=I_{ri}(r)+I_{\vartheta i}(\vartheta),
\end{eqnarray}
where in the last equation we have $i\in\left\{1,2,3\right\}$. In the next step we rewrite the corresponding terms and get
\begin{eqnarray}
&I_{r1}(r)L_{z}^2-2I_{r2}(r)EL_{z}-I_{r3}(r)E^2+I_{\vartheta 1}(\vartheta)L_{z}^2\\
&-2I_{\vartheta 2}(\vartheta)EL_{z}-I_{\vartheta 3}(\vartheta)E^2+H_{1}(r)\left(\frac{\partial S_{r}(r)}{\partial r}\right)^2\nonumber\\
&+H_{2}(\vartheta)\left(\frac{\partial S_{\vartheta}(\vartheta)}{\partial \vartheta}\right)^2+E_{\text{p}r}(r)+E_{\text{p}\vartheta}(\vartheta)=0.\nonumber
\end{eqnarray}
Now we move all $r$-dependent terms to the right-hand side and introduce a separation constant $K$ following the original approach of Carter \cite{Carter1968}. Afterwards we can rewrite the separated equations in the following forms:
\begin{eqnarray}
&H_{1}(r)\left(\frac{\partial S_{r}(r)}{\partial r}\right)^2=I_{r3}(r)E^2+2I_{r2}(r)EL_{z}\\
&-I_{r1}(r)L_{z}^2-E_{\text{p}r}(r)-K,\nonumber
\end{eqnarray} 
\begin{eqnarray}
&H_{2}(\vartheta)\left(\frac{\partial S_{\vartheta}(\vartheta)}{\partial \vartheta}\right)^2=K-E_{\text{p}\vartheta}(\vartheta)+I_{\vartheta 3}(\vartheta)E^2\\
&+2I_{\vartheta 2}(\vartheta)EL_{z}-I_{\vartheta 1}(\vartheta)L_{z}^2.\nonumber
\end{eqnarray}
As we can see the $r$ motion and the $\vartheta$ motion are now completely separated. However, we need the equations of motion in a form that contains the components of the four-velocity and thus we now use (\ref{eq:MAV}) to rewrite them as
\begin{eqnarray}\label{eq:EoMr}
&H_{1}(r)G_{4}(r,\vartheta)^2\dot{r}^2=I_{r3}(r)E^2+2I_{r2}(r)EL_{z}\\
&-I_{r1}(r)L_{z}^2-E_{\text{p}r}(r)-K,\nonumber
\end{eqnarray}
\begin{eqnarray}\label{eq:EoMtheta}
&H_{2}(\vartheta)G_{5}(r,\vartheta)^2\dot{\vartheta}^2=K-E_{\text{p}\vartheta}(\vartheta)+I_{\vartheta 3}(\vartheta)E^2\\
&+2I_{\vartheta 2}(\vartheta)EL_{z}-I_{\vartheta 1}(\vartheta)L_{z}^2.\nonumber
\end{eqnarray}
Now let us also use (\ref{eq:Gi}) to rewrite (\ref{eq:EoMt}) and (\ref{eq:EoMphi}) in a form similar to (\ref{eq:EoMr}) and (\ref{eq:EoMtheta}). When we do this they read
\begin{eqnarray}\label{eq:EoMtr}
&F(r,\vartheta)\dot{t}=I_{r3}(r)E+I_{r2}(r)L_{z}+I_{\vartheta 3}(\vartheta)E\\
&+I_{\vartheta 2}(\vartheta)L_{z},\nonumber
\end{eqnarray}
\begin{eqnarray}\label{eq:EoMphir}
&F(r,\vartheta)\dot{\varphi}=I_{r1}(r)L_{z}-I_{r2}(r)E+I_{\vartheta 1}(\vartheta)L_{z}\\
&-I_{\vartheta 2}(\vartheta)E.\nonumber
\end{eqnarray}
As we can see in this form the equations of motion for $r$ and $\vartheta$ are not completely separated. This is only possible when we can rewrite the left-hand sides of the equations of motion in terms of a Mino parameter $\lambda$ \cite{Mino2003} such that we only have the first derivatives of the spacetime coordinates with respect to the Mino parameter. In addition on the right-hand sides of the equations of motion for $r$ and $\vartheta$ we are only allowed to have terms containing $r$ and $\vartheta$, respectively. Note that in this form the equations of motion for $r$ and $\vartheta$ are separated, however, this does not necessarily imply that they are analytically solvable.

\section{Calculating the Magnification Factor}\label{Sec:MFD}
In the last section we briefly reviewed the separability conditions for the equations of motion of light rays, high-frequency gravitational waves, and massive particles in vacuum as well as light rays in a plasma for stationary and axisymmetric spacetimes. In this section we will now show how we can calculate the magnification factor for a signal emitted by a source and detected by an observer in a stationary and axisymmetric spacetime with separable equations of motion. Here we use the following conventions. When we start at the source the emitted signals move along future-directed trajectories while when we follow a signal from the observer into the past we trace them backwards in time along past-directed trajectories. Now we introduce at the source and the observer an orthonormal tetrad such that the tetrad vectors fulfill the condition
\begin{eqnarray}\label{eq:conv}
g(e_{\mu},e_{\nu})=\eta_{\mu\nu}.
\end{eqnarray}
At the position of the observer we denote the tetrad vectors $e_{O\mu}$, while at the position of the source we denote them $e_{S\nu}$. Note that here $e_{O0}$ and $e_{S0}$ are also the four-velocities of the observer and the source, respectively.
At the position of the observer we can now also write the tangent vector to the trajectory of the signal in terms of the tetrad vectors. It reads
\begin{eqnarray}\label{eq:tanobs}
&\left.\dot{\eta}\right|_{x_{O}}=-\alpha_{O}e_{O0}+\beta_{O}\left(\sin\Sigma_{O}\cos\Psi_{O}e_{O1}\right.\\
&\left.+\sin\Sigma_{O}\sin\Psi_{O}e_{O2}+\cos\Sigma_{O}e_{O3}\right),\nonumber
\end{eqnarray}
where $\Sigma_{O}$ and $\Psi_{O}$ are the latitude and the longitude on the observer's celestial sphere, respectively, and $\alpha_{O}$ and $\beta_{O}$ are normalisation constants. Now using the normalisation condition for the orthonormal vectors (\ref{eq:conv}) it is easy to show that we have 
\begin{eqnarray}
\beta_{O}^2=\alpha_{O}^2-E_{\text{p}}(x_{O})^2.
\end{eqnarray}
After fixing the sign ambiguity of the root such that when we have a static and spherically symmetric spacetime the sign of the angular momentum about the $z$-axis $L_{z}$ agrees between light rays and high-frequency gravitational waves in vacuum and massive particles in vacuum, and light rays in a plasma, we now obtain
\begin{eqnarray}
\beta_{O}=-\sqrt{\alpha_{O}^2-E_{\text{p}}(x_{O})^2}.
\end{eqnarray}
The normalisation constant $\alpha_{O}$ on the other can be calculated from the relation
\begin{eqnarray}
\alpha_{O}=g\left(\left.\dot{\eta}\right|_{x_{O}},e_{O0}\right).
\end{eqnarray}
These relations are generally valid for all signals, however, it is easy to show that in the case of $E_{\text{p}}(x_{O})=0$ we have $\alpha_{O}=\beta_{O}$.

Analogously we can also write the tangent vector to the trajectories at the position of the source as 
\begin{eqnarray}\label{eq:tansrc}
&\left.\dot{\eta}\right|_{x_{S}}=-\left(\alpha_{S}e_{S0}+\beta_{S}\left(\sin\Sigma_{S}\cos\Psi_{S}e_{S1}\right.\right.\\
&\left.\left.+\sin\Sigma_{S}\sin\Psi_{S}e_{S2}+\cos\Sigma_{S}e_{S3}\right)\right),\nonumber
\end{eqnarray}
where this time $\Sigma_{S}$ and $\Psi_{S}$ are the latitude and the longitude on the source's celestial sphere respectively, and $\alpha_{S}$ and $\beta_{S}$ are again normalisation constants. Note that the difference of the sign for the three tetrad vectors $e_{S1}$, $e_{S2}$, and $e_{S3}$ stems from the fact that after being emitted by the source the signals travel along future-directed trajectories. 
Now we again follow the steps outlined above to determine the normalisation constant $\beta_{S}$. The result reads
\begin{eqnarray}
\beta_{S}=-\sqrt{\alpha_{S}^2-E_{\text{p}}(x_{S})^2}.
\end{eqnarray}
The normlisation constant $\alpha_{S}$ on the other hand can be calculated from the relation
\begin{eqnarray}
\alpha_{S}=g\left(\left.\dot{\eta}\right|_{x_{S}},e_{S0}\right).
\end{eqnarray}
Note that again for $E_{\text{p}}(x_{S})=0$ we have $\alpha_{S}=\beta_{S}$. 

In the next step the approach we take differs slightly between light rays and high-frequency gravitational waves travelling in vacuum, and massive particles travelling in vacuum and light rays travelling in a plasma. For the former we can without loss of generality choose $\alpha_{O}=\beta_{O}=-1$ when we start at the observer and $\alpha_{S}=\beta_{S}=-1$ when we start at the source. Together with a comparison of coefficients between the tangent vector to the trajectories in its general form 
\begin{eqnarray}
\dot{\eta}=\dot{t}\partial_{t}+\dot{r}\partial_{r}+\dot{\vartheta}\partial_{\vartheta}+\dot{\varphi}\partial_{\varphi}
\end{eqnarray}
evaluated at the event $x_{O}$ when we start at the observer and at the event $x_{S}$ when we start at the source, and the tangent vectors expressed in terms of the celestial coordinates (\ref{eq:tanobs}) and (\ref{eq:tansrc}), respectively, we can now determine the relations between the constants of motion and the coordinates on the celestial sphere of the observer or the source. 

In the case of massive particles in vacuum and light rays in a plasma we proceed analogously. However, here instead of choosing a value for the normalisation constant we relate the energy of the particles or light rays along the trajectory $E$ to their energy measured either in the reference frame of the observer or the reference frame of the source. They are related to the components of the four-momentum of the signals $p_{\mu}$ and the components of the four-velocities of the observer and the source, respectively, via
\begin{eqnarray}
E_{O}=-p_{\mu}e_{O0}^{\mu}~~~\text{and}~~~E_{S}=-p_{\mu}e_{S0}^{\mu}.
\end{eqnarray}
In the following we will now focus our discussion on a trajectory ending at the position of an observer in the stationary region of the spacetime outside a potentially existing event horizon (commonly also referred to as \emph{domain of outer communication}). 

As described above we first define an orthonormal tetrad at the event at which the observer detects the signal. Then we use the described formalism to relate the constants of motion along the trajectory to the latitude-longitude coordinates on the celestial sphere of the observer. Now the constants of motion are fixed and in the next step we use the same formalism at the position of the source to obtain a relation between the latitude-longitude coordinates on the celestial sphere of the observer and the latitude-longitude coordinates on the celestial sphere of the source. After comparing coefficients we then obtain a system of equations which can be represented in the following form:
\begin{eqnarray}
&f_{1}\left(\Sigma_{O},\Psi_{O},\Sigma_{S},\Psi_{S}\right)=g_{1}\left(\Sigma_{O},\Psi_{O}\right),\label{eq:ME1}\\
&f_{2}\left(\Sigma_{O},\Psi_{O},\Sigma_{S},\Psi_{S}\right)=g_{2}\left(\Sigma_{O},\Psi_{O}\right).\label{eq:ME2}
\end{eqnarray}
Here, the left-hand sides are obtained from the tangent vector rewritten in terms of the tetrad vector and the right-hand sides are obtained from the equations of motion. In addition, the functions on the left-hand side on one hand depend directly on $\Sigma_{O}$ and $\Psi_{O}$ and on the other hand also indirectly through the spacetime coordinates $r_{S}$ and $\vartheta_{S}$ of the source. Note that here the actual form of the dependency of $r_{S}$ and $\vartheta_{S}$ on $\Sigma_{O}$ and $\Psi_{O}$ also depends on the shape of the source surface.

Now we have to distinguish two different cases. In the first case we can solve this system. This allows us to rewrite $\Sigma_{S}$ and $\Psi_{S}$ as functions of $\Sigma_{O}$ and $\Psi_{O}$. In the second case this is generally not possible. Now we have $\Sigma_{S}$ and $\Psi_{S}$ either explicitly or implicitly as functions of $\Sigma_{O}$ and $\Psi_{O}$ and we can use the obtained relations to calculate the magnification factor. For this purpose we need to calculate the determinant of the Jacobi matrix. Let us start with explicitly writing down the Jacobi matrix. In our case it reads
\begin{eqnarray}
J= \left( {\begin{array}{cc}
   \frac{\partial \Sigma_{S}}{\partial \Sigma_{O}} & \frac{\partial \Sigma_{S}}{\partial \Psi_{O}} \\
   \frac{\partial \Psi_{S}}{\partial \Sigma_{O}} & \frac{\partial \Psi_{S}}{\partial \Psi_{O}} \\
  \end{array} } \right).
\end{eqnarray}
Now we have to distinguish the two different cases. In the first case we can express $\Sigma_{S}$ and $\Psi_{S}$ in terms of $\Sigma_{O}$ and $\Psi_{O}$ and thus we can calculate the entries of the Jacobi matrix directly. Then we calculate the determinant of the Jacobi matrix. It reads
\begin{eqnarray}\label{eq:Jac}
\text{det}J=\frac{\partial\Sigma_{S}}{\partial \Sigma_{O}}\frac{\partial \Psi_{S}}{\partial \Psi_{O}}-\frac{\partial \Psi_{S}}{\partial \Sigma_{O}}\frac{\partial \Sigma_{S}}{\partial \Psi_{O}}.
\end{eqnarray}
In our case the determinant characterises the ratio between an infinitesimally small angular area on the celestial sphere of the source and the corresponding infinitesimally small angular area on the celestial sphere of the observer. Now when we assume that the flux through both areas is the same we can use it to determine the magnification factor. It reads
\begin{eqnarray}\label{eq:Mag}
\mu=\frac{1}{\text{det}J}=\frac{1}{\frac{\partial\Sigma_{S}}{\partial \Sigma_{O}}\frac{\partial \Psi_{S}}{\partial \Psi_{O}}-\frac{\partial \Psi_{S}}{\partial \Sigma_{O}}\frac{\partial \Sigma_{S}}{\partial \Psi_{O}}}.
\end{eqnarray}
In the second case we cannot solve for $\Sigma_{S}$ and $\Psi_{S}$. However, we can still calculate the magnification factor. For achieving this we differentiate (\ref{eq:ME1}) and (\ref{eq:ME2}) with respect to $\Sigma_{O}$ and $\Psi_{O}$ and obtain
\begin{eqnarray}
&\frac{\partial f_{1}}{\partial\Sigma_{O}}+\frac{\partial f_{1}}{\partial\Sigma_{S}}\frac{\partial \Sigma_{S}}{\partial\Sigma_{O}}+\frac{\partial f_{1}}{\partial\Psi_{S}}\frac{\partial\Psi_{S}}{\partial\Sigma_{O}}=\frac{\partial g_{1}}{\partial\Sigma_{O}},\\
&\frac{\partial f_{2}}{\partial\Sigma_{O}}+\frac{\partial f_{2}}{\partial\Sigma_{S}}\frac{\partial \Sigma_{S}}{\partial\Sigma_{O}}+\frac{\partial f_{2}}{\partial\Psi_{S}}\frac{\partial\Psi_{S}}{\partial\Sigma_{O}}=\frac{\partial g_{2}}{\partial\Sigma_{O}},\\
&\frac{\partial f_{1}}{\partial\Psi_{O}}+\frac{\partial f_{1}}{\partial\Sigma_{S}}\frac{\partial \Sigma_{S}}{\partial\Psi_{O}}+\frac{\partial f_{1}}{\partial\Psi_{S}}\frac{\partial\Psi_{S}}{\partial\Psi_{O}}=\frac{\partial g_{1}}{\partial\Psi_{O}},\\
&\frac{\partial f_{2}}{\partial\Psi_{O}}+\frac{\partial f_{2}}{\partial\Sigma_{S}}\frac{\partial \Sigma_{S}}{\partial\Psi_{O}}+\frac{\partial f_{2}}{\partial\Psi_{S}}\frac{\partial\Psi_{S}}{\partial\Psi_{O}}=\frac{\partial g_{2}}{\partial\Psi_{O}}.
\end{eqnarray}
We can easily see that we can rewrite these four equations in the form of the following two matrix equations 
\begin{eqnarray}\label{eq:MEOne}
\mathbb{G}\left({\begin{array}{c}
   \frac{\partial \Sigma_{S}}{\partial \Sigma_{O}} \\
   \frac{\partial \Psi_{S}}{\partial \Sigma_{O}} \\
  \end{array} }\right)=\left({\begin{array}{c}
   \frac{\partial g_{1}}{\partial \Sigma_{O}}-\frac{\partial f_{1}}{\partial \Sigma_{O}} \\
   \frac{\partial g_{2}}{\partial \Sigma_{O}}-\frac{\partial f_{2}}{\partial \Sigma_{O}} \\
  \end{array} }\right)
\end{eqnarray}
and 
\begin{eqnarray}\label{eq:METwo}
\mathbb{G}\left({\begin{array}{c}
   \frac{\partial \Sigma_{S}}{\partial \Psi_{O}} \\
   \frac{\partial \Psi_{S}}{\partial \Psi_{O}} \\
  \end{array} }\right)=\left({\begin{array}{c}
   \frac{\partial g_{1}}{\partial \Psi_{O}}-\frac{\partial f_{1}}{\partial \Psi_{O}} \\
   \frac{\partial g_{2}}{\partial \Psi_{O}}-\frac{\partial f_{2}}{\partial \Psi_{O}} \\
  \end{array} }\right),
\end{eqnarray}
where in both equations the matrix $\mathbb{G}$ is given by
\begin{eqnarray}
\mathbb{G}=\left( {\begin{array}{cc}
   \frac{\partial f_{1}}{\partial \Sigma_{S}} & \frac{\partial f_{1}}{\partial \Psi_{S}} \\
   \frac{\partial f_{2}}{\partial \Sigma_{S}} & \frac{\partial f_{2}}{\partial \Psi_{S}} \\
  \end{array} } \right).
\end{eqnarray}
Now when $\text{det}\mathbb{G}\neq 0$ the matrix $\mathbb{G}$ is invertible and its inverse reads
\begin{eqnarray}
\mathbb{G}^{-1}=\frac{1}{\frac{\partial f_{1}}{\partial\Sigma_{S}}\frac{\partial f_{2}}{\partial\Psi_{S}}-\frac{\partial f_{2}}{\partial\Sigma_{S}}\frac{\partial f_{1}}{\partial\Psi_{S}}}\left( {\begin{array}{cc}
   \frac{\partial f_{2}}{\partial \Psi_{S}} & -\frac{\partial f_{1}}{\partial \Psi_{S}} \\
   -\frac{\partial f_{2}}{\partial \Sigma_{S}} & \frac{\partial f_{1}}{\partial \Sigma_{S}} \\
  \end{array} } \right).
\end{eqnarray}
Now we multiply (\ref{eq:MEOne}) and (\ref{eq:METwo}) with $\mathbb{G}^{-1}$ and get 
\begin{eqnarray}
\label{eq:AngDiff1}&\frac{\partial \Sigma_{S}}{\partial\Sigma_{O}}=\frac{\left(\frac{\partial g_{1}}{\partial\Sigma_{O}}-\frac{\partial f_{1}}{\partial\Sigma_{O}}\right)\frac{\partial f_{2}}{\partial\Psi_{S}}-\left(\frac{\partial g_{2}}{\partial\Sigma_{O}}-\frac{\partial f_{2}}{\partial\Sigma_{O}}\right)\frac{\partial f_{1}}{\partial\Psi_{S}}}{\frac{\partial f_{1}}{\partial\Sigma_{S}}\frac{\partial f_{2}}{\partial\Psi_{S}}-\frac{\partial f_{2}}{\partial\Sigma_{S}}\frac{\partial f_{1}}{\partial\Psi_{S}}},\\
\label{eq:AngDiff2}&\frac{\partial \Psi_{S}}{\partial\Sigma_{O}}=\frac{\left(\frac{\partial g_{2}}{\partial\Sigma_{O}}-\frac{\partial f_{2}}{\partial\Sigma_{O}}\right)\frac{\partial f_{1}}{\partial\Sigma_{S}}-\left(\frac{\partial g_{1}}{\partial\Sigma_{O}}-\frac{\partial f_{1}}{\partial\Sigma_{O}}\right)\frac{\partial f_{2}}{\partial\Sigma_{S}}}{\frac{\partial f_{1}}{\partial\Sigma_{S}}\frac{\partial f_{2}}{\partial\Psi_{S}}-\frac{\partial f_{2}}{\partial\Sigma_{S}}\frac{\partial f_{1}}{\partial\Psi_{S}}},\\
\label{eq:AngDiff3}&\frac{\partial \Sigma_{S}}{\partial\Psi_{O}}=\frac{\left(\frac{\partial g_{1}}{\partial\Psi_{O}}-\frac{\partial f_{1}}{\partial\Psi_{O}}\right)\frac{\partial f_{2}}{\partial\Psi_{S}}-\left(\frac{\partial g_{2}}{\partial\Psi_{O}}-\frac{\partial f_{2}}{\partial\Psi_{O}}\right)\frac{\partial f_{1}}{\partial\Psi_{S}}}{\frac{\partial f_{1}}{\partial\Sigma_{S}}\frac{\partial f_{2}}{\partial\Psi_{S}}-\frac{\partial f_{2}}{\partial\Sigma_{S}}\frac{\partial f_{1}}{\partial\Psi_{S}}},\\
\label{eq:AngDiff4}&\frac{\partial \Psi_{S}}{\partial\Psi_{O}}=\frac{\left(\frac{\partial g_{2}}{\partial\Psi_{O}}-\frac{\partial f_{2}}{\partial\Psi_{O}}\right)\frac{\partial f_{1}}{\partial\Sigma_{S}}-\left(\frac{\partial g_{1}}{\partial\Psi_{O}}-\frac{\partial f_{1}}{\partial\Psi_{O}}\right)\frac{\partial f_{2}}{\partial\Sigma_{S}}}{\frac{\partial f_{1}}{\partial\Sigma_{S}}\frac{\partial f_{2}}{\partial\Psi_{S}}-\frac{\partial f_{2}}{\partial\Sigma_{S}}\frac{\partial f_{1}}{\partial\Psi_{S}}}.
\end{eqnarray}
Note that here the right-hand sides generally still depend on $\Sigma_{S}$ and $\Psi_{S}$ and thus we have to solve (\ref{eq:ME1}) and (\ref{eq:ME2}) numerically for given pairs of $\Sigma_{O}$ and $\Psi_{O}$ to calculate the four components of the Jacobi matrix. As final step we now insert (\ref{eq:AngDiff1})--(\ref{eq:AngDiff4}) in (\ref{eq:Mag}) to calculate the magnification factor $\mu$. 

\section{Application to Static and Spherically Symmetric Spacetimes}\label{Sec:ASSSS}
In this section we will apply the introduced formalism to messengers emitted by static sources and detected by a static observer in a static and spherically symmetric spacetime. For this purpose let us first write down the corresponding line element. Using geometric units such that $c=G=1$ it reads
\begin{eqnarray}\label{eq:sssle}
&g_{\mu\nu}\text{d}x^{\mu}\text{d}x^{\nu}=-P_{1}(r)\text{d}t^2+\frac{\text{d}r^2}{P_{2}(r)}+r^2\text{d}\vartheta^2\\
&+r^2\sin^2\vartheta\text{d}\varphi^2.\nonumber
\end{eqnarray}
Here, $P_{1}(r)$ and $P_{2}(r)$ are functions of $r$ which also contain the physical parameters characterising the spacetime. In the most well-known general relativistic static and spherically symmetric spacetimes, such as, e.g., the Schwarzschild spacetime \cite{Schwarzschild1916}, we have $P_{1}(r)=P_{2}(r)$, however, here we keep the line element in its generalised form. Now, following the steps outlined in Section~\ref{Sec:Sep}, we first derive the equations of motion. They read
\begin{eqnarray}\label{eq:EoMtS}
\dot{t}=\frac{E}{P_{1}(r)},
\end{eqnarray}
\begin{eqnarray}\label{eq:EoMrS}
\dot{r}^2=\frac{P_{2}(r)\left(r^2E^2-P_{1}(r)E_{\text{p}r}(r)-P_{1}(r)K\right)}{r^2P_{1}(r)},
\end{eqnarray}
\begin{eqnarray}\label{eq:EoMthetaS}
\dot{\vartheta}^2=\frac{1}{r^4}\left(K-E_{\text{p}\vartheta}(\vartheta)-\frac{L_{z}^2}{\sin^2\vartheta}\right),
\end{eqnarray}
\begin{eqnarray}\label{eq:EoMphiS}
\dot{\varphi}=\frac{L_{z}}{r^2\sin^2\vartheta}.
\end{eqnarray}
Here we have $E_{\text{p}r}(r)=E_{\text{p}\vartheta}(\vartheta)=0$ for light rays and high-frequency gravitational waves travelling along lightlike geodesics in vacuum, note that high-frequency gravitational waves travel along lightlike geodesics was shown by Isaacson \cite{Isaacson1968}, $E_{\text{p}r}(r)=r^2$ and $E_{\text{p}\vartheta}(\vartheta)=0$ for massive particles in vacuum travelling along timelike geodesics, and $E_{\text{p}r}(r)=E_{\text{C}}^2r^2$ and $E_{\text{p}\vartheta}( \vartheta)=0$ for light rays in a homogeneous plasma. In the following we will now illustrate how to apply the method for calculating the magnification factor for the different messenger signals between the reference frames of a static source and a static observer. Note that here we will only demonstrate how to calculate the general structure of the magnification factor. We will not calculate it for specific spacetimes since this will require several lengthy case distinctions which would exceed the scope of this paper. 

\subsection{Light Rays and High-Frequency Gravitational Waves in Vacuum}\label{Sec:LRHFGW}
For light rays and high-frequency gravitational waves travelling along lightlike geodesics in vacuum we have $E_{\text{p}r}(r)=E_{\text{p}\vartheta}(\vartheta)=0$. Thus the equations of motion reduce to
\begin{eqnarray}\label{eq:EoMtLG}
\dot{t}=\frac{E}{P_{1}(r)},
\end{eqnarray}
\begin{eqnarray}\label{eq:EoMrLG}
\dot{r}^2=\frac{P_{2}(r)\left(r^2E^2-P_{1}(r)K\right)}{r^2P_{1}(r)},
\end{eqnarray}
\begin{eqnarray}\label{eq:EoMthetaLG}
\dot{\vartheta}^2=\frac{1}{r^4}\left(K-\frac{L_{z}^2}{\sin^2\vartheta}\right),
\end{eqnarray}
\begin{eqnarray}\label{eq:EoMphiLG}
\dot{\varphi}=\frac{L_{z}}{r^2\sin^2\vartheta}.
\end{eqnarray}
Now let us first specify the local orthonormal tetrad for a static observer detecting a light ray or a high-frequency gravitational wave at the event $x_{O}$ in the line element given by (\ref{eq:sssle}). It reads
\begin{eqnarray}\label{eq:ONT01}
e_{O0}=\left.\frac{\partial_{t}}{\sqrt{P_{1}(r)}}\right|_{x_{O}},~~~e_{O1}=\left.\frac{\partial_{\vartheta}}{r}\right|_{x_{O}},
\end{eqnarray}
\begin{eqnarray}\label{eq:ONT23}
e_{O2}=-\left.\frac{\partial_{\varphi}}{r\sin\vartheta}\right|_{x_{O}},~~~e_{O3}=-\left.\sqrt{P_{2}(r)}\partial_{r}\right|_{x_{O}}.
\end{eqnarray}
In the following we will now, for the convenience of the reader, once step by step illustrate the procedure for deriving the relations between the latitude-longitude coordinates on the celestial sphere of the source and the latitude-longitude coordinates on the celestial sphere of the observer. For this purpose let us again first write down the tangent vector to the lightlike geodesic. It reads
\begin{eqnarray}\label{eq:tanvec}
\dot{\eta}=\dot{t}\partial_{t}+\dot{r}\partial_{r}+\dot{\vartheta}\partial_{\vartheta}+\dot{\varphi}\partial_{\varphi}.
\end{eqnarray}
At the position of the observer we can now also write the tangent vector in terms of the tetrad vectors and the latitude-longitude coordinates $\Sigma_{O}$ and $\Psi_{O}$ on the celestial sphere of the observer. In this case it reads
\begin{eqnarray}\label{eq:tanvecObsLG}
&\left.\dot{\eta}\right|_{x_{O}}=\sigma_{O}\left(-e_{O0}+\sin\Sigma_{O}\cos\Psi_{O}e_{O1}\right.\\
&\left.+\sin\Sigma_{O}\sin\Psi_{O}e_{O2}+\cos\Sigma_{O}e_{O3}\right),\nonumber
\end{eqnarray}
where as already discussed in the last section $\sigma_{O}$ is a normalisation constant. It can be calculated via
\begin{eqnarray}\label{eq:NCLG}
\sigma_{O}=g\left(\left.\dot{\eta}\right|_{x_{O}},e_{O0}\right).
\end{eqnarray}
Here, without loss of generality, we can choose $\sigma_{O}=-1$. When we now use this choice in (\ref{eq:NCLG}) and in (\ref{eq:tanvecObsLG}), and compare coefficients with (\ref{eq:tanvec}) evaluated at the position of the observer at the time when the observer detects the light ray or the high-frequency gravitational wave we obtain for the constants of motion in terms of the celestial coordinates:
\begin{eqnarray}\label{eq:ELG}
E=\sqrt{P_{1}(r_{O})},
\end{eqnarray}
\begin{eqnarray}\label{eq:LzLG}
L_{z}=r_{O}\sin\vartheta_{O}\sin\Sigma_{O}\sin\Psi_{O},
\end{eqnarray}
\begin{eqnarray}\label{eq:KLG}
K=r_{O}^2\sin^2\Sigma_{O}.
\end{eqnarray}
In the next step we write down the orthonormal tetrad for the source. Since we only consider a static source it is simply given by (\ref{eq:ONT01}) and (\ref{eq:ONT23}) but we have to evaluate the tetrad vectors at $x_{S}$ instead of $x_{O}$. We also write down the tangent vector to the light ray at the position of the source. Since here it is future-directed it reads in terms of the tetrad vetcors and the latitude-longitude coordinates on the celestial sphere of the source $\Sigma_{S}$ and $\Psi_{S}$ 
\begin{eqnarray}\label{eq:tanvecSrcLG}
&\left.\dot{\eta}\right|_{x_{S}}=-\sigma_{S}\left(e_{S0}+\sin\Sigma_{S}\cos\Psi_{S}e_{S1}\right.\\
&\left.+\sin\Sigma_{S}\sin\Psi_{S}e_{S2}+\cos\Sigma_{S}e_{S3}\right),\nonumber
\end{eqnarray}
where again $\sigma_{S}$ is a normalisation constant. However, since the energy of the signal is already fixed in this case we cannot choose it arbitrarily. Using the same procedure as above we obtain that it is given by
\begin{eqnarray}
\sigma_{S}=g\left(\left.\dot{\eta}\right|_{x_{S}},e_{S0}\right)=-\sqrt{\frac{P_{1}(r_{O})}{P_{1}(r_{S})}}.
\end{eqnarray}
We insert $\sigma_{S}$ in (\ref{eq:tanvecSrcLG}) and compare coefficients with (\ref{eq:tanvec}) evaluated at the event $x_{S}$ when the source emits the light ray or the high-frequency gravitational wave. As a result we obtain two relations. The first reads 
\begin{eqnarray}\label{eq:LGR1}
&\frac{r_{O}\sin\vartheta_{O}\sin\Sigma_{O}\sin\Psi_{O}}{\sqrt{P_{1}(r_{O})}}=-\frac{r_{S}\sin\vartheta_{S}\sin\Sigma_{S}\sin\Psi_{S}}{\sqrt{P_{1}(r_{S})}}.
\end{eqnarray}
From the second relation we directly obtain a relation between the latitude on the celestial sphere of the observer $\Sigma_{O}$ and the latitude on the celestial sphere of the source $\Sigma_{S}$. Here we have to distinguish two different cases. They read
\begin{widetext}
\begin{eqnarray}\label{eq:VacSig}
\Sigma_{S}=
\begin{cases}
\arcsin\left(\frac{r_{O}}{r_{S}}\sqrt{\frac{P_{1}(r_{S})}{P_{1}(r_{O})}}\sin\Sigma_{O}\right)~~~~~~~~\text{for}~~~0\leq\Sigma_{S}\leq \frac{\pi}{2},\\
\pi-\arcsin\left(\frac{r_{O}}{r_{S}}\sqrt{\frac{P_{1}(r_{S})}{P_{1}(r_{O})}}\sin\Sigma_{O}\right)~~~\text{for}~~~\frac{\pi}{2}<\Sigma_{S}\leq \pi.\\
\end{cases}
\end{eqnarray}
\end{widetext}
Note that here in general $\sin\Sigma_{O}$ is not unique and thus we have to choose the specific case at hand based on the direction of the motion, in our case the direction of the $r$ motion, of the light ray at the position of the source. However, fortunately for past- and future-directed motion along the same geodesic the first derivatives have the same sign and thus conveniently we can make this choice based on the direction of motion between the last turning point, or the observer if there is no turning point, and the source.

Now we insert the obtained result in (\ref{eq:LGR1}) and solve for $\Psi_{S}$ to obtain a relation between the longitude on the celestial sphere of the observer $\Psi_{O}$ and the longitude on the celestial sphere of the source $\Psi_{S}$. This time we have to distinguish three different cases and get
\begin{widetext}
\begin{eqnarray}\label{eq:VacPsi}
\Psi_{S}=
\begin{cases}
-\arcsin\left(\frac{\sin\vartheta_{O}\sin\Psi_{O}}{\sin\vartheta_{S}}\right)~~~~~~~~\text{for}~~~0\leq\Psi_{S}\leq \frac{\pi}{2},\\
\pi+\arcsin\left(\frac{\sin\vartheta_{O}\sin\Psi_{O}}{\sin\vartheta_{S}}\right)~~~~~\text{for}~~~\frac{\pi}{2}<\Psi_{S}\leq \frac{3\pi}{2},\\
2\pi-\arcsin\left(\frac{\sin\vartheta_{O}\sin\Psi_{O}}{\sin\vartheta_{S}}\right)~~~~\text{for}~~~\frac{3\pi}{2}<\Psi_{S}.\\
\end{cases}
\end{eqnarray}
\end{widetext}
Also here in the cases in which $\sin\Psi_{O}$ is not unique we have to choose the correct case based on the direction of motion, in our case the direction of the $\vartheta$ motion, between the last turning point (or in the case that there is no turning point along the trajectory the observer) and the source. 

In the next step we need to calculate the components of the Jacobi matrix. For this purpose we now calculate the derivatives of $\Sigma_{S}$ and $\Psi_{S}$ with respect to $\Sigma_{O}$ and $\Psi_{O}$. Here we assume two different scenarios. In the first scenario we assume that we have sources distributed on a two-sphere at the radius coordinate $r_{S}$. In the second scenario we assume that we have sources distributed in a luminous disk in the equatorial plane.

\subsubsection{Sources on a Two-Sphere}
In this case we have light and gravitational wave sources distributed on the surface of a two-sphere. Therefore, we have $r_{S}=\text{const.}$ and $\vartheta_{S}=\vartheta_{S}(\Sigma_{O},\Psi_{O})$. We first calculate the derivatives of $\Sigma_{S}$ with respect to $\Sigma_{O}$. As we can easily read from (\ref{eq:VacSig}) the derivatives of the different cases with respect to $\Sigma_{O}$ are identical up to a sign and read
\begin{eqnarray}\label{eq:dVacSigSig}
&\frac{\partial\Sigma_{S}}{\partial\Sigma_{O}}=\pm r_{O}\cos\Sigma_{O}\sqrt{\frac{P_{1}(r_{S})}{r_{S}^2P_{1}(r_{O})-r_{O}^2P_{1}(r_{S})\sin^2\Sigma_{O}}},
\end{eqnarray}
where the plus sign has to be chosen for $0\leq\Sigma_{S}\leq\pi/2$ and the minus sign has to be chosen for $\pi/2<\Sigma_{S}\leq\pi$. Since the right-hand side of (\ref{eq:VacSig}) does not depend on $\Psi_{O}$ in this case we get for the derivative of $\Sigma_{S}$ with respect to $\Psi_{O}$:
\begin{eqnarray}
\frac{\partial\Sigma_{S}}{\partial\Psi_{O}}=0.
\end{eqnarray} 
In the next step we calculate the derivatives of (\ref{eq:VacPsi}) with respect to $\Sigma_{O}$ and $\Psi_{O}$. Again it is easy to see that the derivatives of the different cases on the right-hand side of (\ref{eq:VacPsi}) only differ up to a sign. We first calculate the derivative with respect to $\Sigma_{O}$. It reads
\begin{eqnarray}\label{eq:dVacPsiSig}
\frac{\partial\Psi_{S}}{\partial\Sigma_{O}}=\pm\frac{\sin\vartheta_{O}\cot\vartheta_{S}\sin\Psi_{O}\partial_{\Sigma_{O}}\vartheta_{S}}{\sqrt{\sin^2\vartheta_{S}-\sin^2\vartheta_{O}\sin^2\Psi_{O}}},
\end{eqnarray}
where the plus sign has to be chosen for $0\leq\Psi_{S}\leq\pi/2$ and $3\pi/2<\Psi_{S}$ and the minus sign has to be chosen for $\pi/2<\Psi_{S}\leq3\pi/2$. In the next step we calculate the derivative with respect to $\Psi_{O}$. It reads
\begin{eqnarray}\label{eq:dVacPsiPsi}
\frac{\partial\Psi_{S}}{\partial\Psi_{O}}=
\pm\frac{\sin\vartheta_{O}\left(\cot\vartheta_{S}\sin\Psi_{O}\partial_{\Psi_{O}}\vartheta_{S}-\cos\Psi_{O}\right)}{\sqrt{\sin^2\vartheta_{S}-\sin^2\vartheta_{O}\sin^2\Psi_{O}}},
\end{eqnarray}
where again the plus sign has to be chosen for $0\leq\Psi_{S}\leq\pi/2$ and $3\pi/2<\Psi_{S}$ and the minus sign has to be chosen for $\pi/2<\Psi_{S}\leq3\pi/2$.
Now we calculate the determinant of the Jacobi matrix and the magnification factor. Since we have $\partial_{\Psi_{O}}\Sigma_{S}=0$ the second term of the determinant in (\ref{eq:Jac}) vanishes and thus in this case the magnifcation factor reads
\begin{widetext}
\begin{eqnarray}\label{eq:magvac}
\mu=\pm\frac{\sqrt{\sin^2\vartheta_{S}-\sin^2\vartheta_{O}\sin^2\Psi_{O}}}{r_{O}\cos\Sigma_{O}\sin\vartheta_{O}\left(\sin\Psi_{O}\cot\vartheta_{S}\partial_{\Psi_{O}}\vartheta_{S}-\cos\Psi_{O}\right)}\sqrt{\frac{r_{S}^2P_{1}(r_{O})-r_{O}^2P_{1}(r_{S})\sin^2\Sigma_{O}}{P_{1}(r_{S})}},
\end{eqnarray}
\end{widetext}
where the plus sign has to be chosen for $0\leq\Sigma_{S}\leq\pi/2$ and $0\leq \Psi_{S}\leq \pi/2$ or $3\pi/2<\Psi_{S}$, or $\pi/2<\Sigma_{S}\leq\pi$ and $\pi/2<\Psi_{S}\leq 3\pi/2$, and the minus sign has to be chosen for $0\leq\Sigma_{S}\leq\pi/2$ and $\pi/2<\Psi_{S}\leq 3\pi/2$, or $\pi/2<\Sigma_{S}\leq\pi$ and $0\leq \Psi_{S}\leq \pi/2$ or $3\pi/2<\Psi_{S}$.

\subsubsection{Sources in the Equatorial Plane}\label{Sec:EP}
In the second case we have sources distributed in the equatorial plane. Since here we only want to derive the basic relations for calculating the magnification factor we do not really need to specify the boundaries of the disk, however, when one wants to calculate the magnification factor for real astrophysical scenarios, e.g., an accretion disk, the disk will usually be limited by an inner boundary $r_{\text{in}}$ and an outer boundary $r_{\text{out}}$. 

In this case we have $r_{S}=r_{S}(\Sigma_{O},\Psi_{O})$ and $\vartheta_{S}=\pi/2=\text{const}.$ Note that for the explicit calculations we keep $\vartheta_{S}$ in the equations so that the relations can be easily transferred to other scenarios. The equatorial case can be easily obtained by setting $\vartheta_{S}=\pi/2$ throughout the calculations. Again we start with calculating the derivatives of $\Sigma_{S}$ with respect to $\Sigma_{O}$ and $\Psi_{O}$. In this case the derivative of $\Sigma_{S}$ with respect to $\Sigma_{O}$ reads
\begin{widetext}
\begin{eqnarray}
\frac{\partial\Sigma_{S}}{\partial \Sigma_{O}}=\pm\frac{r_{O}\left(2r_{S}P_{1}(r_{S})\cos\Sigma_{O}+\sin\Sigma_{O}\left(r_{S} \partial_{r_{S}} P_{1}(r_{S})-2P_{1}(r_{S})\right)\partial_{\Sigma_{O}}r_{S}\right)}{2r_{S}\sqrt{P_{1}(r_{S})\left(r_{S}^2P_{1}(r_{O})-r_{O}^2P_{1}(r_{S})\sin^2\Sigma_{O}\right)}},
\end{eqnarray}
\end{widetext}
where we have to choose the plus sign for $0\leq \Sigma_{S}\leq \pi/2$ and the minus sign for $\pi/2<\Sigma_{S}\leq \pi$. Similarly for the derivative with respect to $\Psi_{O}$ we get

\begin{eqnarray}
&\frac{\partial\Sigma_{S}}{\partial \Psi_{O}}=\pm\frac{r_{O}\sin\Sigma_{O}\left(r_{S} \partial_{r_{S}} P_{1}(r_{S})-2P_{1}(r_{S})\right)\partial_{\Psi_{O}}r_{S}}{2r_{S}\sqrt{P_{1}(r_{S})\left(r_{S}^2P_{1}(r_{O})-r_{O}^2P_{1}(r_{S})\sin^2\Sigma_{O}\right)}},
\end{eqnarray}
where again we have to choose the plus sign for $0\leq \Sigma_{S}\leq \pi/2$ and the minus sign for $\pi/2<\Sigma_{S}\leq \pi$. In the next step we calculate the derivative of $\Psi_{S}$ with respect to $\Sigma_{O}$. Since in this case we have $\vartheta_{S}=\text{const}.$ it is easy to see that we get
\begin{eqnarray}\label{eq:PsiSSigmaO}
\frac{\partial\Psi_{S}}{\partial \Sigma_{O}}=0.
\end{eqnarray}
The derivative of $\Psi_{S}$ with respect to $\Psi_{O}$ on the other hand reads
\begin{eqnarray}\label{eq:PsiSPsiO}
\frac{\partial\Psi_{S}}{\partial \Psi_{O}}=\mp\frac{\sin\vartheta_{O}\cos\Psi_{O}}{\sqrt{\sin^2\vartheta_{S}-\sin^2\vartheta_{O}\sin^2\Psi_{O}}},
\end{eqnarray}
where we have to choose the minus sign for $0\leq \Psi_{S}\leq\pi/2$ and $3\pi/2<\Psi_{S}$ and the plus sign for $\pi/2<\Psi_{S}\leq 3\pi/2$. Finally, we again calculate the determinant of the Jacobi matrix. It is easy to see that again the second term of the determinant in (\ref{eq:Jac}) vanishes since we have $\partial_{\Sigma_{O}}\Psi_{S}=0$. As a consequence the magnification factor reads
\begin{widetext}
\begin{eqnarray}\label{eq:magvacep}
\mu=\pm\frac{2r_{S}\sqrt{P_{1}(r_{S})\left(r_{S}^2P_{1}(r_{O})-r_{O}^2P_{1}(r_{S})\sin^2\Sigma_{O}\right)\left(\sin^2\vartheta_{S}-\sin^2\vartheta_{O}\sin^2\Psi_{O}\right)}}{r_{O}\sin\vartheta_{O}\cos\Psi_{O}\left(\left(2P_{1}(r_{S})-r_{S}\partial_{r_{S}}P_{1}(r_{S})\right)\sin\Sigma_{O}\partial_{\Sigma_{O}}r_{S}-2r_{S}P_{1}(r_{S})\cos\Sigma_{O}\right)},
\end{eqnarray}
\end{widetext}
where again we have to choose the plus sign for $0\leq\Sigma_{S}\leq\pi/2$ and $0\leq \Psi_{S}\leq \pi/2$ or $3\pi/2<\Psi_{S}$, or $\pi/2<\Sigma_{S}\leq\pi$ and $\pi/2<\Psi_{S}\leq 3\pi/2$, and the minus sign for $0\leq\Sigma_{S}\leq\pi/2$ and $\pi/2<\Psi_{S}\leq 3\pi/2$, or $\pi/2<\Sigma_{S}\leq\pi$ and $0\leq \Psi_{S}\leq \pi/2$ or $3\pi/2<\Psi_{S}$.

\subsection{Massive Particles in Vacuum and Light Rays in a Homogeneous Plasma}
Now we turn to the case of massive particles in vacuum and light rays in a homogeneous plasma. Since we only consider a static and spherically symmetric line element in both cases we have $E_{\text{p}\vartheta}(\vartheta)=0$. For the $r$-dependent parts we have $E_{\text{p}r}(r)=r^2$ for massive particles in vacuum and $E_{\text{p}r}(r)=E_{\text{C}}^2 r^2$ for light rays in a homogeneous plasma. We can easily see that the function $E_{\text{p}r}(r)$ for massive particles in vacuum is a special case of the function $E_{\text{p}r}(r)$ for light rays in a homogeneous plasma and thus in the following we will write all equations in this generalised form. The specific results for massive particles in vacuum can then simply be obtained by setting $E_{\text{C}}=1$. Now we again first follow the steps outlined in Section~\ref{Sec:MFD}. This time we relate the constants of motion to the energy of the signals measured in the reference frame of the observer $E_{O}$ and the latitude-longitude coordinates on the observer's celestial sphere $\Sigma_{O}$ and $\Psi_{O}$. This time the obtained relations read:
\begin{eqnarray}
E=\sqrt{P_{1}(r_{O})}E_{O},
\end{eqnarray}
\begin{eqnarray}
L_{z}=\sqrt{E_{O}^2-E_{\text{C}}^2}r_{O}\sin\vartheta_{O}\sin\Sigma_{O}\sin\Psi_{O},
\end{eqnarray}
\begin{eqnarray}
K=\left(E_{O}^2-E_{\text{C}}^2\right)r_{O}^2\sin^2\Sigma_{O}.
\end{eqnarray}
As we can see this time the constants of motion all depend on the energy of the massive particle or the light ray measured at the position of the observer $E_{O}$. 

In the next step we again use the same procedure at the position of the source to relate the latitude-longitude coordinates in the reference frame of the source $\Sigma_{S}$ and $\Psi_{S}$ to the latitude-longitude coordinates in the reference frame of the observer $\Sigma_{O}$ and $\Psi_{O}$. This time we obtain for the relation between the celestial latitude in the reference frame of the source $\Sigma_{S}$ and the celestial latitude in the reference frame of the observer $\Sigma_{O}$:
\begin{widetext}
\begin{eqnarray}\label{eq:VacSigP}
\Sigma_{S}=
\begin{cases}
\arcsin\left(\frac{r_{O}}{r_{S}}\sqrt{\frac{P_{1}(r_{S})(E_{O}^2-E_{\text{C}}^2)}{P_{1}(r_{O})E_{O}^2-P_{1}(r_{S})E_{\text{C}}^2}}\sin\Sigma_{O}\right)~~~~~~~~\text{for}~~~0\leq\Sigma_{S}\leq \frac{\pi}{2},\\
\pi-\arcsin\left(\frac{r_{O}}{r_{S}}\sqrt{\frac{P_{1}(r_{S})(E_{O}^2-E_{\text{C}}^2)}{P_{1}(r_{O})E_{O}^2-P_{1}(r_{S})E_{\text{C}}^2}}\sin\Sigma_{O}\right)~~~\text{for}~~~\frac{\pi}{2}<\Sigma_{S}\leq \pi.\\
\end{cases}
\end{eqnarray}
\end{widetext}
Again $\sin\Sigma_{O}$ does not necessarily allow to uniquely determine which case we have to choose and thus we have to make this decision based on the direction of the $r$ motion at the position of the source.

The relation between the celestial longitude in the reference frame of the source $\Psi_{S}$ and the celestial longitude in the reference frame of the observer $\Psi_{O}$ on the other hand is again given by (\ref{eq:VacPsi}). 

Now we again assume the same two different cases as for light rays and high-frequency gravitational waves in vacuum, namely sources on a two-sphere at the radius coordinate $r_{S}$ and sources in a luminous disk in the equatorial plane. As in Section~\ref{Sec:LRHFGW} we will first derive the entries of the Jacobi matrix and then the magnification factor $\mu$.

\subsubsection{Sources on a Two-Sphere}
As for light rays and high-frequency gravitational waves in vacuum we again begin with sources distributed on a two-sphere. We again have $r_{S}=\text{const}.$ and $\vartheta_{S}=\vartheta_{S}\left(\Sigma_{O},\Psi_{O}\right)$. We follow the same steps as described above to obtain the entries of the Jacobi matrix. We start with deriving the derivatives of $\Sigma_{S}$ with respect to $\Sigma_{O}$ and $\Psi_{O}$. In the first case we get 
\begin{widetext}
\begin{eqnarray}\label{eq:VacSigMPL}
\frac{\partial\Sigma_{S}}{\partial\Sigma_{O}}=\pm r_{O}\cos\Sigma_{O}\sqrt{\frac{P_{1}(r_{S})\left(E_{O}^2-E_{\text{C}}^2\right)}{r_{S}^2\left(P_{1}(r_{O})E_{O}^2-P_{1}(r_{S})E_{\text{C}}^2\right)-r_{O}^2P_{1}(r_{s})\left(E_{O}^2-E_{\text{C}}^2\right)\sin^2\Sigma_{O}}},
\end{eqnarray}
\end{widetext}
where we have to choose the plus sign for $0\leq\Sigma_{S}\leq\pi/2$ and the minus sign for $\pi/2<\Sigma_{S}\leq \pi$. In the second case on the other hand we get
\begin{eqnarray}
\frac{\partial\Sigma_{S}}{\partial\Psi_{O}}=0.
\end{eqnarray}
Now we still need the derivatives of $\Psi_{S}$ with respect to $\Sigma_{O}$ and $\Psi_{O}$. Since $\Psi_{S}$ is related to $\Psi_{O}$ by (\ref{eq:VacPsi}) the derivatives are again given by (\ref{eq:dVacPsiSig}) and (\ref{eq:dVacPsiPsi}). In the next step we again calculate the determinant of the Jacobi matrix and finally the magnification factor $\mu$. In this case the magnification factor reads
\begin{widetext}
\begin{eqnarray}
&\mu=\pm\frac{\sqrt{\sin^2\vartheta_{S}-\sin^2\vartheta_{O}\sin^2\Psi_{O}}}{r_{O}\cos\Sigma_{O}\sin\vartheta_{O}\left(\sin\Psi_{O}\cot\vartheta_{S}\partial_{\Psi_{O}}\vartheta_{S}-\cos\Psi_{O}\right)}\sqrt{\frac{r_{S}^2\left(P_{1}(r_{O})E_{O}^2-P_{1}(r_{S})E_{\text{C}}^2\right)-r_{O}^2P_{1}(r_{S})(E_{O}^2-E_{\text{C}}^2)\sin^2\Sigma_{O}}{P_{1}(r_{S})(E_{O}^2-E_{\text{C}}^2)}},
\end{eqnarray}
\end{widetext}
where we again have to choose the plus sign for $0\leq\Sigma_{S}\leq\pi/2$ and $0\leq \Psi_{S}\leq \pi/2$ or $3\pi/2<\Psi_{S}$, or $\pi/2<\Sigma_{S}\leq\pi$ and $\pi/2<\Psi_{S}\leq 3\pi/2$, and the minus sign for $0\leq\Sigma_{S}\leq\pi/2$ and $\pi/2<\Psi_{S}\leq 3\pi/2$, or $\pi/2<\Sigma_{S}\leq\pi$ and $0\leq \Psi_{S}\leq \pi/2$ or $3\pi/2<\Psi_{S}$. 

When we compare the obtained magnification factor to its counterpart for light rays and high-frequency gravitational waves given by (\ref{eq:magvac}) we can easily see that the main differences are that for massive particles in vacuum and light rays in a homogeneous plasma the obtained expression for the magnification factor is slightly more complex and depends on the energy of the particle or the light ray measured by the observer. In addition, it is also easy to see that when we set $E_{\text{C}}=0$ the magnification factor for massive particles in vacuum and light rays in a homogeneous plasma reduces to the magnification factor for light rays and high-frequency gravitational waves in vacuum.

\subsubsection{Sources in the Equatorial Plane}
As for light rays and high-frequency gravitational waves in vacuum in the second case we have sources in a luminous disk in the equatorial plane. We again have $r_{S}=r_{S}(\Sigma_{O},\Psi_{O})$ and $\vartheta_{S}=\pi/2=\text{const}.$ As in Section~\ref{Sec:EP} we keep all equations in their general form and the specific case for sources in the equatorial plane can be obtained by setting $\vartheta_{S}=\pi/2$. Again we first derive the entries of the Jacobi matrix. In this case the derivative of the celestial latitude in the reference frame of the source $\Sigma_{S}$ with respect to the celestial latitude in the reference frame of the observer $\Sigma_{O}$ reads
\begin{widetext}
\begin{eqnarray}
&\frac{\partial\Sigma_{S}}{\partial \Sigma_{O}}=\pm\sqrt{\frac{P_{1}(r_{S})(E_{O}^2-E_{\text{C}}^2)}{r_{S}^2\left(P_{1}(r_{O})E_{O}^2-P_{1}(r_{S})E_{\text{C}}^2\right)-r_{O}^2P_{1}(r_{S})(E_{O}^2-E_{\text{C}}^2)\sin^2\Sigma_{O}}}\\
&\frac{r_{O}\left(\left(r_{S}P_{1}(r_{O})E_{O}^2\partial_{r_{S}}P_{1}(r_{S})-2P_{1}(r_{S})\left(P_{1}(r_{O})E_{O}^2-P_{1}(r_{S})E_{\text{C}}^2\right)\right)\sin\Sigma_{O}\partial_{\Sigma_{O}}r_{S}+2P_{1}(r_{S})r_{S}\left(P_{1}(r_{O})E_{O}^2-P_{1}(r_{S})E_{\text{C}}^2\right)\cos\Sigma_{O}\right)}{2r_{S}P_{1}(r_{S})\left(P_{1}(r_{O})E_{O}^2-P_{1}(r_{S})E_{\text{C}}^2\right)},\nonumber
\end{eqnarray}
\end{widetext}
where the plus sign has to be chosen for $0\leq\Sigma_{S}\leq\pi/2$ while the minus sign has to be chosen for $\pi/2<\Sigma_{S}\leq\pi$. The derivative of the celestial latitude in the reference frame of the source $\Sigma_{S}$ with respect to the celestial longitude in the reference frame of the observer $\Psi_{O}$ on the other hand reads
\begin{widetext}
\begin{eqnarray}
&\frac{\partial\Sigma_{S}}{\partial \Psi_{O}}=\pm\sqrt{\frac{P_{1}(r_{S})(E_{O}^2-E_{\text{C}}^2)}{r_{S}^2\left(P_{1}(r_{O})E_{O}^2-P_{1}(r_{S})E_{\text{C}}^2\right)-r_{O}^2P_{1}(r_{S})(E_{O}^2-E_{\text{C}}^2)\sin^2\Sigma_{O}}}\\
&\frac{r_{O}\sin\Sigma_{O}\left(r_{S}P_{1}(r_{O})E_{O}^2\partial_{r_{S}}P_{1}(r_{S})-2P_{1}(r_{S})\left(P_{1}(r_{O})E_{O}^2-P_{1}(r_{S})E_{\text{C}}^2\right)\right)\partial_{\Psi_{O}}r_{S}}{2r_{S}P_{1}(r_{S})\left(P_{1}(r_{O})E_{O}^2-P_{1}(r_{S})E_{\text{C}}^2\right)},\nonumber
\end{eqnarray}
\end{widetext}
where again the plus sign has to be chosen for $0\leq\Sigma_{S}\leq\pi/2$ while the minus sign has to be chosen for $\pi/2<\Sigma_{S}\leq\pi$. Now we still have to calculate the derivatives of the celestial longitude in the reference frame of the source $\Psi_{S}$ with respect to the latitude-longitude coordinates $\Sigma_{O}$ and $\Psi_{O}$ in the reference frame of the observer. As already mentioned above in this case the relation between $\Psi_{S}$ and $\Sigma_{O}$ and $\Psi_{O}$ is the same as for light rays and high-frequency gravitational waves in vacuum and thus also in this case these derivatives are given by (\ref{eq:PsiSSigmaO}) and (\ref{eq:PsiSPsiO}), respectively. In the next step we again use the derived entries of the Jacobi matrix to derive its determinant. Then we use the determinant of the Jacobi matrix to calculate the magnification factor for massive particles in vacuum (for $E_{\text{C}}=1$) and light rays in a homogeneous plasma. It reads
\onecolumngrid
\begin{eqnarray}
&\mu=\pm\sqrt{\frac{r_{S}^2\left(P_{1}(r_{O})E_{O}^2-P_{1}(r_{S})E_{\text{C}}^2\right)-r_{O}^2P_{1}(r_{S})(E_{O}^2-E_{\text{C}}^2)\sin^2\Sigma_{O}}{P_{1}(r_{S})(E_{O}^2-E_{\text{C}}^2)}}\\
&\frac{2r_{S}P_{1}(r_{S})\left(P_{1}(r_{O})E_{O}^2-P_{1}(r_{S})E_{\text{C}}^2\right)\sqrt{\sin^2\vartheta_{S}-\sin^2\vartheta_{O}\sin^2\Psi_{O}}}{r_{O}\sin\vartheta_{O}\cos\Psi_{O}\left(\left(2P_{1}(r_{S})\left(P_{1}(r_{O})E_{O}^2-P_{1}(r_{S})E_{\text{C}}^2\right)-r_{S}P_{1}(r_{O})E_{O}^2\partial_{r_{S}}P_{1}(r_{S})\right)\sin\Sigma_{O}\partial_{\Sigma_{O}}r_{S}+2r_{S}P_{1}(r_{S})\left(P_{1}(r_{S})E_{\text{C}}^2-P_{1}(r_{O})E_{O}^2\right)\cos\Sigma_{O}\right)},\nonumber
\end{eqnarray}
\twocolumngrid
where we again have to choose the plus sign for $0\leq\Sigma_{S}\leq\pi/2$ and $0\leq \Psi_{S}\leq \pi/2$ or $3\pi/2<\Psi_{S}$, or $\pi/2<\Sigma_{S}\leq\pi$ and $\pi/2<\Psi_{S}\leq 3\pi/2$, and the minus sign for $0\leq\Sigma_{S}\leq\pi/2$ and $\pi/2<\Psi_{S}\leq 3\pi/2$, or $\pi/2<\Sigma_{S}\leq\pi$ and $0\leq \Psi_{S}\leq \pi/2$ or $3\pi/2<\Psi_{S}$.

Again when we compare the obtained expression for the magnification factor for massive particles and light rays in a homogeneous plasma with the corresponding expression for the magnification factor for light rays and high-frequency gravitational waves in vacuum (\ref{eq:magvacep}) we see that it is slightly more complex and depends on the energy of the massive particle or the light ray at the position of the observer. In addition, we also see that when we set $E_{\text{C}}=0$ the magnification factor for massive particles and light rays in a homogeneous plasma reduces to the magnification factor for light rays and high-frequency gravitational waves in vacuum (\ref{eq:magvacep}).

\section{Summary}\label{Sec:SO}
In this paper we showed that for stationary and axisymmetric spacetimes with separable equations of motion, namely motion for which we have a Carter constant, we can derive a magnification factor which relates the flux through an infinitesimally small angular area element on the celestial sphere of a source to the flux through an infinitesimally small angular area element on the celestial sphere of an observer. For this purpose we first reviewed under which conditions the equations of motion of stationary and axisymmetric spacetimes can be separated, following the work of Bezd\v{e}kov\'{a}, Perlick, and Bi\v{c}\'{a}k \cite{Bezdekova2022}. 

In the next step we used that when we introduce local orthonormal tetrads at the positions of the source and the observer, we can write down a set of equations that relates the latitude-longitude coordinates on the celestial sphere of the source to the latitude-longitude coordinates on the celestial sphere of the observer. Finally, we used this set of equations to derive the entries of the Jacobi matrix, its determinant, and the magnification factor. While in this paper we demonstrated the basic steps assuming that we start at the position of the observer this approach can also be easily transferred to situations in which we start at the source.

In the most general case we cannot solve the derived set of equations for the latitude-longitude coordinates on, in our case, the celestial sphere of the source. In addition, also the Jacobi matrix and thus also the magnification factor still contain the latitude-longitude coordinates on the celestial sphere of the source. Thus in general for determining the corresponding latitude-longitude coordinates in the reference frame of the source and the magnification factor, we have to solve this system of equations numerically and then insert the obtained values in the magnification factor. However, in special cases we can also solve the equations for the latitude-longitude coordinates on the celestial sphere of the source. This enables us to calculate the magnification factor purely analytically when the equations of motion for $r$ and $\vartheta$ possess analytic solutions.

For demonstrating the usefulness of the proposed method in the last part of this paper we applied it to derive the magnification factor for light rays, high-frequency gravitational waves, and massive particles in vacuum, and light rays in a homogeneous plasma in a general static and spherically symmetric spacetime. Here we assumed two different scenarios. In the first we considered light sources distributed on the surface of a two-sphere. In the second scenario we considered a luminous disk in the equatorial plane. 

In both scenarios we obtained magnification factors which contain the physical parameters of the spacetime, the spacetime coordinates of the observer, and the latitude-longitude coordinates on the celestial sphere of the observer. In addition, for massive particles in vacuum and light rays in a homogeneous plasma the magnification factor also contained the energy measured at the position of the observer. Furthermore, all magnification factors also contained the derivative of the spacetime latitude (for sources on the surface of a two-sphere) or of the radius coordinate (for sources in a luminous disk in the equatorial plane) of the source with respect to one of the coordinates on the celestial sphere of the observer. Here, we now have two options to evaluate the magnification factors. On one hand we can calculate the derivatives using numerical approximation methods or, when the equations of motion admit analytical solutions, we can derive them analytically. Examples for spacetimes for which this is possible are, e.g., the Schwarzschild spacetime or the Kerr spacetime. However, in the case that we have the solutions to the equations of motion in analytical form calculating the magnification factor requires several case distinctions, which can easily get quite complicated, see, e.g., the work of Frost \cite{Frost2024} for the equations of motion in the Kerr spacetime. Thus this was left for future work. 

Here, it will be particularly interesting to see which of the most commonly chosen orthonormal tetrads for stationary and axisymmetric spacetimes, namely the tetrad of an observer (or source) on a $t$-line, the tetrad of a zero angular momentum observer \cite{Bardeen1972}, and the tetrad of a standard observer \cite{Grenzebach2015} allow to derive the magnification factor purely analytically.

The last point we have to address is now which advantages the presented method has compared to already existing methods. First of all, as already pointed out in the introduction, the most common methods either use the lens equation to derive the magnification factor with respect to the spacetime position of the source or the affine and shape parameters. Second, these magnification factors usually relate the size of or the flux or the intensity associated with a source and measured by an observer in a curved spacetime to the same quantities of the source measured by an observer in a flat spacetime. While this allows to investigate the spacetime structure of the lens it does not \emph{a-priori} allow to investigate the properties of the source. 

The magnification factor derived in this paper now allows to relate the flux through an infinitesimally small angular area element on the celestial sphere of an observer to the flux through the associated infinitesimally small angular area element on the celestial sphere of a source. When we detect an image on our sky or a signal in a detector we can measure the former. In addition, when we can use the signal or independent measurements to determine the spacetime describing the lens, and we can infer the motion of the source and the observer relative to the lens, the calculated magnification factor will provide us with direct access to the flux in the reference frame of the source. This will then allow us to determine the emission properties of the source and to investigate the source's physical properties.

Here, the most likely candidates for applying the presented method are supermassive black hole candidates. They are usually described by stationary and axisymmetric spacetimes and thus the presented approach may prove useful for investigating emission structures in the accretion disk close to them when higher-resolution images become available. While right now the resolution of the Event Horizon Telescope is only high enough to resolve the shadows of and the rough structure of the accretion disks around the supermassive black hole candidates in the centres of the galaxy M87 \cite{EHTCollaboration2019a} and the Milky Way \cite{EHTCollaboration2022}, the next generation Event Horizon Telescope \cite{Johnson2023} and the Black Hole Explorer \cite{Johnson2024} will allow to observe the central regions of both galaxies at a higher resolution and this may also allow to observe more details in their accretion disks.
\section*{Acknowledments}
I would like to thank Che-Yu Chen and Xian Chen for their comments and our discussions. I acknowledge funding from the China Postdoctoral Science Foundation (Grant No. 2023M740111) and from the National Key Research and Development Program of China (Grant No. 2021YFC2203002).
\bibliography{Magnification.bib}

\begin{thebibliography}{20}%
\makeatletter
\providecommand \@ifxundefined [1]{%
 \@ifx{#1\undefined}
}%
\providecommand \@ifnum [1]{%
 \ifnum #1\expandafter \@firstoftwo
 \else \expandafter \@secondoftwo
 \fi
}%
\providecommand \@ifx [1]{%
 \ifx #1\expandafter \@firstoftwo
 \else \expandafter \@secondoftwo
 \fi
}%
\providecommand \natexlab [1]{#1}%
\providecommand \enquote  [1]{``#1''}%
\providecommand \bibnamefont  [1]{#1}%
\providecommand \bibfnamefont [1]{#1}%
\providecommand \citenamefont [1]{#1}%
\providecommand \href@noop [0]{\@secondoftwo}%
\providecommand \href [0]{\begingroup \@sanitize@url \@href}%
\providecommand \@href[1]{\@@startlink{#1}\@@href}%
\providecommand \@@href[1]{\endgroup#1\@@endlink}%
\providecommand \@sanitize@url [0]{\catcode `\\12\catcode `\$12\catcode
  `\&12\catcode `\#12\catcode `\^12\catcode `\_12\catcode `\%12\relax}%
\providecommand \@@startlink[1]{}%
\providecommand \@@endlink[0]{}%
\providecommand \url  [0]{\begingroup\@sanitize@url \@url }%
\providecommand \@url [1]{\endgroup\@href {#1}{\urlprefix }}%
\providecommand \urlprefix  [0]{URL }%
\providecommand \Eprint [0]{\href }%
\providecommand \doibase [0]{https://doi.org/}%
\providecommand \selectlanguage [0]{\@gobble}%
\providecommand \bibinfo  [0]{\@secondoftwo}%
\providecommand \bibfield  [0]{\@secondoftwo}%
\providecommand \translation [1]{[#1]}%
\providecommand \BibitemOpen [0]{}%
\providecommand \bibitemStop [0]{}%
\providecommand \bibitemNoStop [0]{.\EOS\space}%
\providecommand \EOS [0]{\spacefactor3000\relax}%
\providecommand \BibitemShut  [1]{\csname bibitem#1\endcsname}%
\let\auto@bib@innerbib\@empty
%</preamble>
\bibitem [{\citenamefont {Perlick}(2004{\natexlab{a}})}]{Perlick2004b}%
  \BibitemOpen
  \bibfield  {author} {\bibinfo {author} {\bibfnamefont {V.}~\bibnamefont
  {Perlick}},\ }\bibfield  {title} {\bibinfo {title} {Gravitational lensing
  from a spacetime perspective},\ }\href {https://doi.org/10.12942/lrr-2004-9}
  {\bibfield  {journal} {\bibinfo  {journal} {Living {R}ev. {R}elativ.}\
  }\textbf {\bibinfo {volume} {7}},\ \bibinfo {pages} {9} (\bibinfo {year}
  {2004}{\natexlab{a}})}\BibitemShut {NoStop}%
\bibitem [{\citenamefont {Petters}\ \emph {et~al.}(2001)\citenamefont
  {Petters}, \citenamefont {Levine},\ and\ \citenamefont
  {Wambsganss}}]{Petters2001}%
  \BibitemOpen
  \bibfield  {author} {\bibinfo {author} {\bibfnamefont {A.~O.}\ \bibnamefont
  {Petters}}, \bibinfo {author} {\bibfnamefont {H.}~\bibnamefont {Levine}},\
  and\ \bibinfo {author} {\bibfnamefont {J.}~\bibnamefont {Wambsganss}},\
  }\href {https://doi.org/10.1007/978-1-4612-0145-8} {\emph {\bibinfo {title}
  {Singularity Theory and Gravitational Lensing}}},\ \bibinfo {edition} {1st}\
  ed.,\ Progress in {M}athematical {P}hysics\ (\bibinfo  {publisher}
  {Birkhäuser Boston, MA},\ \bibinfo {year} {2001})\BibitemShut {NoStop}%
\bibitem [{\citenamefont {Frittelli}\ and\ \citenamefont
  {Newman}(1999)}]{Frittelli1999}%
  \BibitemOpen
  \bibfield  {author} {\bibinfo {author} {\bibfnamefont {S.}~\bibnamefont
  {Frittelli}}\ and\ \bibinfo {author} {\bibfnamefont {E.~T.}\ \bibnamefont
  {Newman}},\ }\bibfield  {title} {\bibinfo {title} {Exact universal
  gravitational lensing equation},\ }\href
  {https://doi.org/10.1103/PhysRevD.59.124001} {\bibfield  {journal} {\bibinfo
  {journal} {Phys. {R}ev. {D}}\ }\textbf {\bibinfo {volume} {59}},\ \bibinfo
  {pages} {124001} (\bibinfo {year} {1999})}\BibitemShut {NoStop}%
\bibitem [{\citenamefont {Frittelli}\ \emph
  {et~al.}(2001{\natexlab{a}})\citenamefont {Frittelli}, \citenamefont
  {Kling},\ and\ \citenamefont {Newman}}]{Frittelli2001a}%
  \BibitemOpen
  \bibfield  {author} {\bibinfo {author} {\bibfnamefont {S.}~\bibnamefont
  {Frittelli}}, \bibinfo {author} {\bibfnamefont {T.~P.}\ \bibnamefont
  {Kling}},\ and\ \bibinfo {author} {\bibfnamefont {E.~T.}\ \bibnamefont
  {Newman}},\ }\bibfield  {title} {\bibinfo {title} {Image distortion in
  nonperturbative gravitational lensing},\ }\href
  {https://doi.org/10.1103/PhysRevD.63.023006} {\bibfield  {journal} {\bibinfo
  {journal} {Phys. {R}ev. {D}}\ }\textbf {\bibinfo {volume} {63}},\ \bibinfo
  {pages} {023006} (\bibinfo {year} {2001}{\natexlab{a}})}\BibitemShut
  {NoStop}%
\bibitem [{\citenamefont {Frittelli}\ \emph
  {et~al.}(2001{\natexlab{b}})\citenamefont {Frittelli}, \citenamefont
  {Kling},\ and\ \citenamefont {Newman}}]{Frittelli2001b}%
  \BibitemOpen
  \bibfield  {author} {\bibinfo {author} {\bibfnamefont {S.}~\bibnamefont
  {Frittelli}}, \bibinfo {author} {\bibfnamefont {T.~P.}\ \bibnamefont
  {Kling}},\ and\ \bibinfo {author} {\bibfnamefont {E.~T.}\ \bibnamefont
  {Newman}},\ }\bibfield  {title} {\bibinfo {title} {Image distortion from
  optical scalars in nonperturbative gravitational lensing},\ }\href
  {https://doi.org/10.1103/PhysRevD.63.023007} {\bibfield  {journal} {\bibinfo
  {journal} {Phys. {R}ev. {D}}\ }\textbf {\bibinfo {volume} {63}},\ \bibinfo
  {pages} {023007} (\bibinfo {year} {2001}{\natexlab{b}})}\BibitemShut
  {NoStop}%
\bibitem [{\citenamefont {Kraniotis}(2011)}]{Kraniotis2011}%
  \BibitemOpen
  \bibfield  {author} {\bibinfo {author} {\bibfnamefont {G.~V.}\ \bibnamefont
  {Kraniotis}},\ }\bibfield  {title} {\bibinfo {title} {Precise analytic
  treatment of {K}err and {K}err-(anti) de {S}itter black holes as
  gravitational lenses},\ }\href
  {https://doi.org/10.1088/0264-9381/28/8/085021} {\bibfield  {journal}
  {\bibinfo  {journal} {Classical {Q}uantum {G}ravity}\ }\textbf {\bibinfo
  {volume} {28}},\ \bibinfo {pages} {085021} (\bibinfo {year}
  {2011})}\BibitemShut {NoStop}%
\bibitem [{\citenamefont {Perlick}(2004{\natexlab{b}})}]{Perlick2004a}%
  \BibitemOpen
  \bibfield  {author} {\bibinfo {author} {\bibfnamefont {V.}~\bibnamefont
  {Perlick}},\ }\bibfield  {title} {\bibinfo {title} {Exact gravitational lens
  equation in spherically symmetric and static spacetimes},\ }\href
  {https://doi.org/10.1103/PhysRevD.69.064017} {\bibfield  {journal} {\bibinfo
  {journal} {Phys. {R}ev. {D}}\ }\textbf {\bibinfo {volume} {69}},\ \bibinfo
  {pages} {064017} (\bibinfo {year} {2004}{\natexlab{b}})}\BibitemShut
  {NoStop}%
\bibitem [{\citenamefont {Bezd\v{e}kov\'{a}}\ \emph {et~al.}(2022)\citenamefont
  {Bezd\v{e}kov\'{a}}, \citenamefont {Perlick},\ and\ \citenamefont
  {Bi\v{c}\'{a}k}}]{Bezdekova2022}%
  \BibitemOpen
  \bibfield  {author} {\bibinfo {author} {\bibfnamefont {B.}~\bibnamefont
  {Bezd\v{e}kov\'{a}}}, \bibinfo {author} {\bibfnamefont {V.}~\bibnamefont
  {Perlick}},\ and\ \bibinfo {author} {\bibfnamefont {J.}~\bibnamefont
  {Bi\v{c}\'{a}k}},\ }\bibfield  {title} {\bibinfo {title} {Light propagation
  in a plasma on an axially symmetric and stationary spacetime: {S}eparability
  of the {H}amilton--{J}acobi equation and shadow},\ }\href
  {https://doi.org/10.1063/5.0106433} {\bibfield  {journal} {\bibinfo
  {journal} {J. {M}ath. {P}hys.}\ }\textbf {\bibinfo {volume} {63}},\ \bibinfo
  {pages} {092501} (\bibinfo {year} {2022})}\BibitemShut {NoStop}%
\bibitem [{\citenamefont {Carter}(1968)}]{Carter1968}%
  \BibitemOpen
  \bibfield  {author} {\bibinfo {author} {\bibfnamefont {B.}~\bibnamefont
  {Carter}},\ }\bibfield  {title} {\bibinfo {title} {Global structure of the
  {K}err family of gravitational fields},\ }\href
  {https://doi.org/10.1103/PhysRev.174.1559} {\bibfield  {journal} {\bibinfo
  {journal} {Phys. {R}ev.}\ }\textbf {\bibinfo {volume} {174}},\ \bibinfo
  {pages} {1559} (\bibinfo {year} {1968})}\BibitemShut {NoStop}%
\bibitem [{\citenamefont {Perlick}\ and\ \citenamefont
  {Tsupko}(2017)}]{Perlick2017}%
  \BibitemOpen
  \bibfield  {author} {\bibinfo {author} {\bibfnamefont {V.}~\bibnamefont
  {Perlick}}\ and\ \bibinfo {author} {\bibfnamefont {O.~Y.}\ \bibnamefont
  {Tsupko}},\ }\bibfield  {title} {\bibinfo {title} {Light propagation in a
  plasma on {K}err spacetime: {S}eparation of the {H}amilton-{J}acobi equation
  and calculation of the shadow},\ }\href
  {https://doi.org/10.1103/PhysRevD.95.104003} {\bibfield  {journal} {\bibinfo
  {journal} {Phys. {R}ev. {D}}\ }\textbf {\bibinfo {volume} {95}},\ \bibinfo
  {pages} {104003} (\bibinfo {year} {2017})}\BibitemShut {NoStop}%
\bibitem [{\citenamefont {Mino}(2003)}]{Mino2003}%
  \BibitemOpen
  \bibfield  {author} {\bibinfo {author} {\bibfnamefont {Y.}~\bibnamefont
  {Mino}},\ }\bibfield  {title} {\bibinfo {title} {Perturbative approach to an
  orbital evolution around a supermassive black hole},\ }\href
  {https://doi.org/10.1103/PhysRevD.67.084027} {\bibfield  {journal} {\bibinfo
  {journal} {Phys. {R}ev. {D}}\ }\textbf {\bibinfo {volume} {67}},\ \bibinfo
  {pages} {084027} (\bibinfo {year} {2003})}\BibitemShut {NoStop}%
\bibitem [{\citenamefont {Schwarzschild}(1916)}]{Schwarzschild1916}%
  \BibitemOpen
  \bibfield  {author} {\bibinfo {author} {\bibfnamefont {K.}~\bibnamefont
  {Schwarzschild}},\ }\bibfield  {title} {\bibinfo {title} {\"{U}ber das
  {G}ravitationsfeld eines {M}assenpunktes nach der {E}insteinschen
  {T}heorie},\ }\href@noop {} {\bibfield  {journal} {\bibinfo  {journal}
  {Sitzungsberichte der {K}\"{o}niglich-{P}reu{\ss}ischen {A}kademie der
  {W}issenschaften}\ }\textbf {\bibinfo {volume} {VII}},\ \bibinfo {pages}
  {189} (\bibinfo {year} {1916})}\BibitemShut {NoStop}%
\bibitem [{\citenamefont {Isaacson}(1968)}]{Isaacson1968}%
  \BibitemOpen
  \bibfield  {author} {\bibinfo {author} {\bibfnamefont {R.~A.}\ \bibnamefont
  {Isaacson}},\ }\bibfield  {title} {\bibinfo {title} {Gravitational radiation
  in the limit of high frequency. {I}. {T}he linear approximation and
  geometrical optics},\ }\href {https://doi.org/10.1103/PhysRev.166.1263}
  {\bibfield  {journal} {\bibinfo  {journal} {Phys. {R}ev.}\ }\textbf {\bibinfo
  {volume} {166}},\ \bibinfo {pages} {1263} (\bibinfo {year}
  {1968})}\BibitemShut {NoStop}%
\bibitem [{\citenamefont {Frost}(2024)}]{Frost2024}%
  \BibitemOpen
  \bibfield  {author} {\bibinfo {author} {\bibfnamefont {T.~C.}\ \bibnamefont
  {Frost}},\ }\href {https://doi.org/10.48550/arXiv.2411.13368} {\bibinfo
  {title} {Gravitational lensing in the {K}err spacetime: An analytic approach
  for light and high-frequency gravitational waves}},\ \bibinfo {howpublished}
  {arXiv:2411.13368,} (\bibinfo {year} {2024})\BibitemShut {NoStop}%
\bibitem [{\citenamefont {Bardeen}\ \emph {et~al.}(1972)\citenamefont
  {Bardeen}, \citenamefont {Press},\ and\ \citenamefont
  {Teukolsky}}]{Bardeen1972}%
  \BibitemOpen
  \bibfield  {author} {\bibinfo {author} {\bibfnamefont {J.~M.}\ \bibnamefont
  {Bardeen}}, \bibinfo {author} {\bibfnamefont {W.~H.}\ \bibnamefont {Press}},\
  and\ \bibinfo {author} {\bibfnamefont {S.~A.}\ \bibnamefont {Teukolsky}},\
  }\bibfield  {title} {\bibinfo {title} {Rotating black holes: {L}ocally
  nonrotating frames, energy extraction, and scalar synchrotron radiation},\
  }\href {https://doi.org/10.1086/151796} {\bibfield  {journal} {\bibinfo
  {journal} {{A}strophys. {J}.}\ }\textbf {\bibinfo {volume} {178}},\ \bibinfo
  {pages} {347} (\bibinfo {year} {1972})}\BibitemShut {NoStop}%
\bibitem [{\citenamefont {Grenzebach}\ \emph {et~al.}(2015)\citenamefont
  {Grenzebach}, \citenamefont {Perlick},\ and\ \citenamefont
  {L\"ammerzahl}}]{Grenzebach2015}%
  \BibitemOpen
  \bibfield  {author} {\bibinfo {author} {\bibfnamefont {A.}~\bibnamefont
  {Grenzebach}}, \bibinfo {author} {\bibfnamefont {V.}~\bibnamefont
  {Perlick}},\ and\ \bibinfo {author} {\bibfnamefont {C.}~\bibnamefont
  {L\"ammerzahl}},\ }\bibfield  {title} {\bibinfo {title} {Photon regions and
  shadows of accelerated black holes},\ }\href
  {https://doi.org/10.1142/S0218271815420249} {\bibfield  {journal} {\bibinfo
  {journal} {Int. {J}. {M}od. {P}hys. {D}}\ }\textbf {\bibinfo {volume} {24}},\
  \bibinfo {pages} {1542024} (\bibinfo {year} {2015})}\BibitemShut {NoStop}%
\bibitem [{\citenamefont {{The Event Horizon Telescope Collaboration}}\ \emph
  {et~al.}(2019)\citenamefont {{The Event Horizon Telescope Collaboration}}
  \emph {et~al.}}]{EHTCollaboration2019a}%
  \BibitemOpen
  \bibfield  {author} {\bibinfo {author} {\bibnamefont {{The Event Horizon
  Telescope Collaboration}}} \emph {et~al.},\ }\bibfield  {title} {\bibinfo
  {title} {First {M}87 {E}vent {H}orizon {T}elescope results. {I}. {T}he shadow
  of the supermassive black hole},\ }\href
  {https://doi.org/10.3847/2041-8213/ab0ec7} {\bibfield  {journal} {\bibinfo
  {journal} {{A}strophys. {J}. {L}ett.}\ }\textbf {\bibinfo {volume} {875}},\
  \bibinfo {pages} {L1} (\bibinfo {year} {2019})}\BibitemShut {NoStop}%
\bibitem [{\citenamefont {{The Event Horizon Telescope Collaboration}}\ \emph
  {et~al.}(2022)\citenamefont {{The Event Horizon Telescope Collaboration}}
  \emph {et~al.}}]{EHTCollaboration2022}%
  \BibitemOpen
  \bibfield  {author} {\bibinfo {author} {\bibnamefont {{The Event Horizon
  Telescope Collaboration}}} \emph {et~al.},\ }\bibfield  {title} {\bibinfo
  {title} {First {S}agittarius $\text{A}^{*}$ {E}vent {H}orizon {T}elescope
  results. {I}. {T}he shadow of the supermassive black hole in the center of
  the {M}ilky {W}ay},\ }\href {https://doi.org/10.3847/2041-8213/ac6674}
  {\bibfield  {journal} {\bibinfo  {journal} {{A}strophys. {J}. {L}ett.}\
  }\textbf {\bibinfo {volume} {930}},\ \bibinfo {pages} {L12} (\bibinfo {year}
  {2022})}\BibitemShut {NoStop}%
\bibitem [{\citenamefont {Johnson}\ \emph {et~al.}(2023)\citenamefont {Johnson}
  \emph {et~al.}}]{Johnson2023}%
  \BibitemOpen
  \bibfield  {author} {\bibinfo {author} {\bibfnamefont {M.~D.}\ \bibnamefont
  {Johnson}} \emph {et~al.},\ }\bibfield  {title} {\bibinfo {title} {Key
  science goals for the next-generation {E}vent {H}orizon {T}elescope},\ }\href
  {https://doi.org/10.3390/galaxies11030061} {\bibfield  {journal} {\bibinfo
  {journal} {Galaxies}\ }\textbf {\bibinfo {volume} {11}},\ \bibinfo {pages}
  {61} (\bibinfo {year} {2023})}\BibitemShut {NoStop}%
\bibitem [{\citenamefont {Johnson}\ \emph {et~al.}(2024)\citenamefont {Johnson}
  \emph {et~al.}}]{Johnson2024}%
  \BibitemOpen
  \bibfield  {author} {\bibinfo {author} {\bibfnamefont {M.~D.}\ \bibnamefont
  {Johnson}} \emph {et~al.},\ }\bibfield  {title} {\bibinfo {title} {The
  {B}lack {H}ole {E}xplorer: motivation and vision},\ }in\ \href
  {https://doi.org/10.1117/12.3019835} {\emph {\bibinfo {booktitle} {Space
  {T}elescopes and {I}nstrumentation 2024: {O}ptical, {I}nfrared, and
  {M}illimeter {W}ave}}},\ Vol.\ \bibinfo {volume} {13092},\ \bibinfo {editor}
  {edited by\ \bibinfo {editor} {\bibfnamefont {L.~E.}\ \bibnamefont {Coyle}},
  \bibinfo {editor} {\bibfnamefont {S.}~\bibnamefont {Matsuura}},\ and\
  \bibinfo {editor} {\bibfnamefont {M.~D.}\ \bibnamefont {Perrin}}},\ \bibinfo
  {organization} {International Society for Optics and Photonics}\ (\bibinfo
  {publisher} {SPIE},\ \bibinfo {year} {2024})\ p.\ \bibinfo {pages}
  {130922D}\BibitemShut {NoStop}%
\end{thebibliography}%

\end{document}